\documentclass[preprint,showpacs,12pt]{revtex4}
\usepackage{amsmath,amssymb,amsfonts}
\usepackage{graphicx}
\usepackage{hhline}
\usepackage{bm}
\usepackage{amsmath}
\usepackage{epsfig}

\begin{document}
\begin{flushleft}
CERN-PH-TH/2012-046\\
\end{flushleft}
\title{Neutrino energy reconstruction problems and neutrino oscillations}
\author {M. Martini}
\affiliation{Institut d'Astronomie et d'Astrophysique, CP-226, Universit\'e Libre de Bruxelles, 1050 Brussels, Belgium}
\author {M. Ericson}
\affiliation{Universit\'e de Lyon, Univ. Lyon 1,
 CNRS/IN2P3, IPN Lyon, F-69622 Villeurbanne Cedex, France}
\affiliation{Physics Department, Theory Unit, CERN, CH-1211 Geneva, Switzerland}
\author {G. Chanfray}
\affiliation{Universit\'e de Lyon, Univ. Lyon 1,
 CNRS/IN2P3, IPN Lyon, F-69622 Villeurbanne Cedex, France}

\begin{abstract}

 We discuss the accuracy of the usual procedure for neutrino energy reconstruction which is based on the quasielastic kinematics.  Our results are described in terms of a probability distribution for a real neutrino energy value. Several factors  are responsible for the deviations from the reconstructed value. The main one is the multinucleon component of the neutrino interaction which in the case of Cherenkov detectors enters as a quasielastic cross section,  increasing the mean neutrino energy which can differ appreciably from the reconstructed value.
   
 As an application we derive, for excess electron events attributed to the conversion of muon neutrinos, the true neutrino energy distribution based on the experimental one which is given in terms of the reconstructed value. The result is a reshaping effect. For MiniBooNE  the low energy peak is suppressed and shifted at higher energies, which may influence the interpretation in terms of oscillation. For T2K at the Super Kamiokande far detector the reshaping translates into a narrowing of the energy distribution.
\end{abstract}

\pacs{13.15.+g, 25.30.Pt, 14.60.Pq}
\maketitle


\section{Introduction}     
Neutrino oscillation experiments require the determination of the neutrino energy which enters the expression of the oscillation probability. 
This determination is commonly done through the charged current quasielastic events, commonly defined as those in which the emission product 
only includes one charged lepton, the ejected nucleon being unobservable.
 
 For the ``quasielastic'' events where only the charged lepton is observed the only measurable quantities are then its direction, i.e., its emission angle $\theta$ with respect to the neutrino beam direction and its energy $E_l$ (or kinetic energy $T_l$ and momentum $P_l$).  The neutrino energy  $E_\nu$ is unknown.  
The usual reconstruction procedure assumes that we are dealing with a genuine quasielastic event on a nucleon at rest. In this case the 
energy $\omega$ and momentum $q$ transferred to the nuclear system are related by:  $\omega = (q^2- \omega^2)/(2M)$, where $M$ is the nucleon mass. 
These quantities are related to the charged lepton observables by: $\omega=E_\nu- E_l$ and 
$ q^2 = E^2_\nu +  P^2_l - 2  E_\nu P_l \cos{\theta} $. For illustration we take the example of an ejected muon. 
The quasielastic condition then gives the value $\overline{E_\nu}$  of the reconstructed energy
\begin{equation}
\label{eq1_erec}
\overline{E_\nu} = {E_\mu-m^2_\mu/(2M)\over1-(E_\mu-P_\mu \cos \theta)/ M}.
\end {equation}
A binding correction can be introduced in this expression but it is irrelevant for our discussion.
Several effects can influence this expression. First the Fermi motion which broadens the quasielastic peak, the Pauli blocking  which cuts part of the low momentum response, but also the fact that a number of events such as the multinucleon ones which are not genuine quasielastic ones can, in the case for instance of Cherenkov detectors, simulate those but have no reason to fulfill the quasielastic relation. The multinucleon component  has been shown 
\cite{Martini:2009uj,Martini:2010ex,Martini:2011wp,Nieves:2011pp,Nieves:2011yp}
to be responsible for a sizeable increase of the quasielastic cross section. 

\begin{figure}
\begin{center}
  \includegraphics[width=16cm,height=12cm]{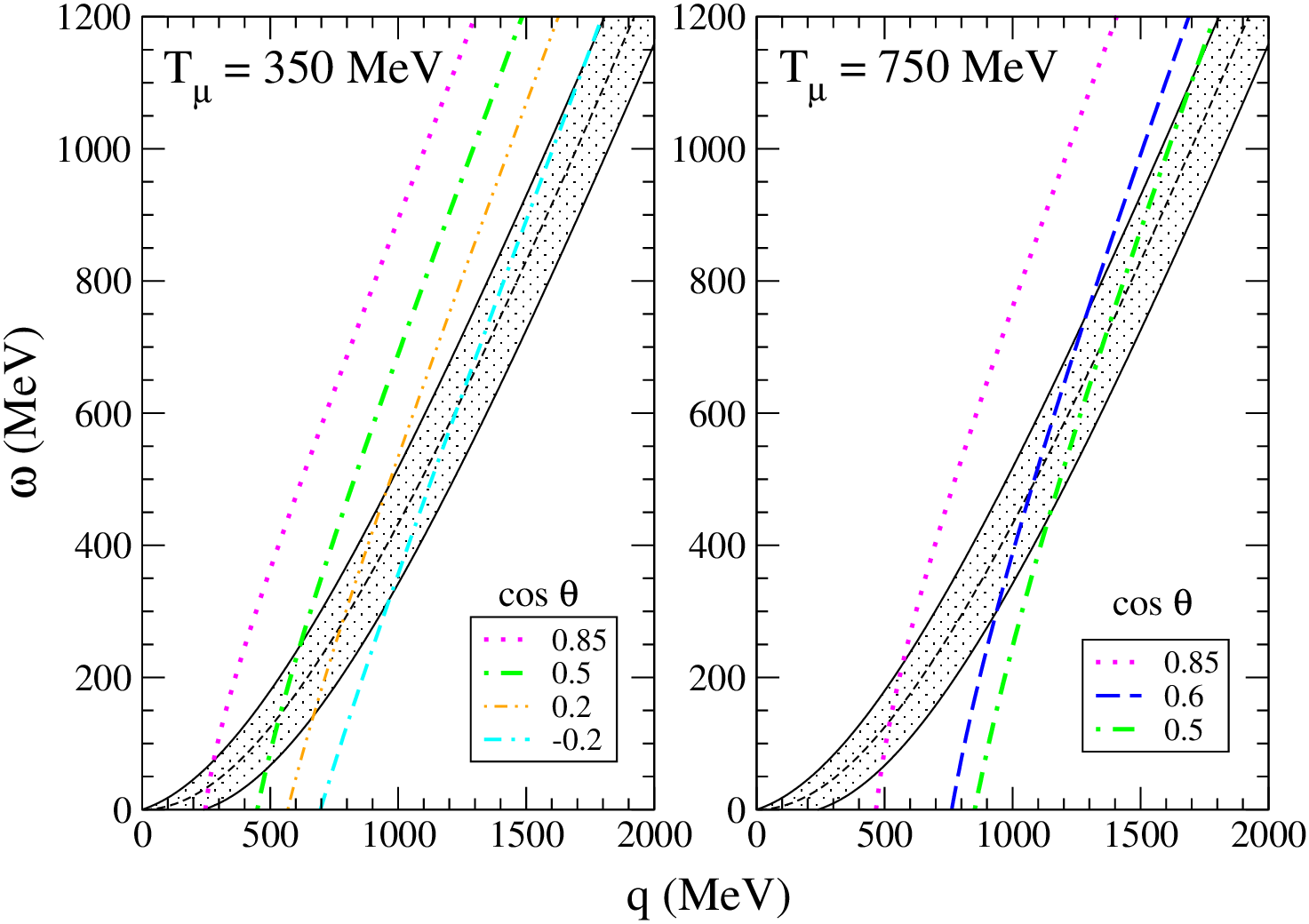}
\caption{(Color online) The neutrino 
hyperbolas defined by Eq. (\ref{eq_hyp}) for a muon kinetic energy $T_\mu$=350 MeV, $T_\mu$=750 MeV and several muon emission angles indicated in the figure.
The shaded area delimited by the two corresponding continuous lines represents the region of the quasielastic response of a Fermi gas. The central dashed lines show the position of the quasielastic peak.}
\label{fig_hyperbolas}
\end{center}
\end{figure}

In order to visualize the various effects on the reconstruction of the neutrino energy we first repeat an argument of Refs. 
\cite{Delorme:1985ps} and \cite{Martini:2011wp}. Eliminating $E_\nu=\omega+ E_\mu $ in the expression of $q^2$ leads to the relation 
\begin{equation}
\label{eq_hyp}
q^2 - \omega^2= 4 (E_\mu+\omega) E_\mu ~\mathrm{sin} ^2 \frac{\theta}{2}-m_{\mu}^2+2 (E_\mu+\omega) (E_\mu-p_\mu) \cos {\theta}.
\end{equation} 
In the $\omega$ and  $q$ plane this equation represents a series of hyperbolas, which depend on the values of $E_\mu$ and 
$\theta$, with asymptotes parallel
to the $\omega = q $ line.  Examples are shown in Fig. \ref{fig_hyperbolas}. The intercept of the hyperbola with the region of response  of the nucleus, whatever its origin, quasielastic or not,  fixes the possible $\omega$ and $q$ values for these  values of $E_\mu$
 and $\theta$. In the case of a dilute Fermi gas where the region of response reduces to the quasielastic line, the  intercept values $\omega =\omega_{\textrm{intercept}}$ and $q= q_{\textrm{intercept}}$ are completely fixed. Hence the neutrino energy is also determined : 
$E_\nu = E_\mu + \omega_{\textrm{intercept}}$, which leads to the usual value of the reconstructed energy for this  set of muon observables, $E_\mu$ and $\theta$. 
The Fermi motion introduces a broadening around the quasielastic line, as represented by the shaded area of Fig. \ref{fig_hyperbolas}. We will show that in some situations this broadening is not innocent for the reconstruction problem. In addition the quasielastic peak itself can be distorted by Pauli correlations or by the collective nature of the response, as described in the random phase approximation (RPA) \cite{Martini:2009uj}. These effects destroy the symmetry of the distribution of the $E_\nu$ values around the  reconstructed energy  $\overline{E_\nu}$. 
Moreover in a realistic description with  correlations between nucleons, more than one nucleon can be ejected by the neutrino  interaction. 
The Delta resonance can also be excited in this process with subsequent emission of several nucleons. 
This has the effect of enlarging the region of response to the whole  $\omega$ and $q$ plane. 
This means that for a given set of muon variables, $E_\mu$ and $\theta$, an infinity of  $\omega$ values is possible, hence of neutrino energies, 
instead of the unique quasielastic value implemented in the neutrino energy reconstruction formula. 
The question is to find the distribution of the possible  $ E_\nu $ values and to understand the influence of all these factors on the energy reconstruction. 
The results are described in terms of probability distributions to have a true neutrino energy between $E_\nu$ 
and $E_\nu+dE_\nu$. 
Prior investigations by Benhar and Meloni  \cite{Benhar:2009wi} have dealt with the influence of the nuclear spectral function. 
Another work by Leitner and Mosel \cite{Leitner:2010kp} 
deals with the question of real pions produced and then absorbed on their way out of the nucleus, treated by transport equation. 
This process as well simulates a quasielastic interaction. In our work instead we omit the final state interaction, hence the absorption of real pions produced.
 
\begin{figure}
\begin{center}
\includegraphics[width=12cm,height=8cm]{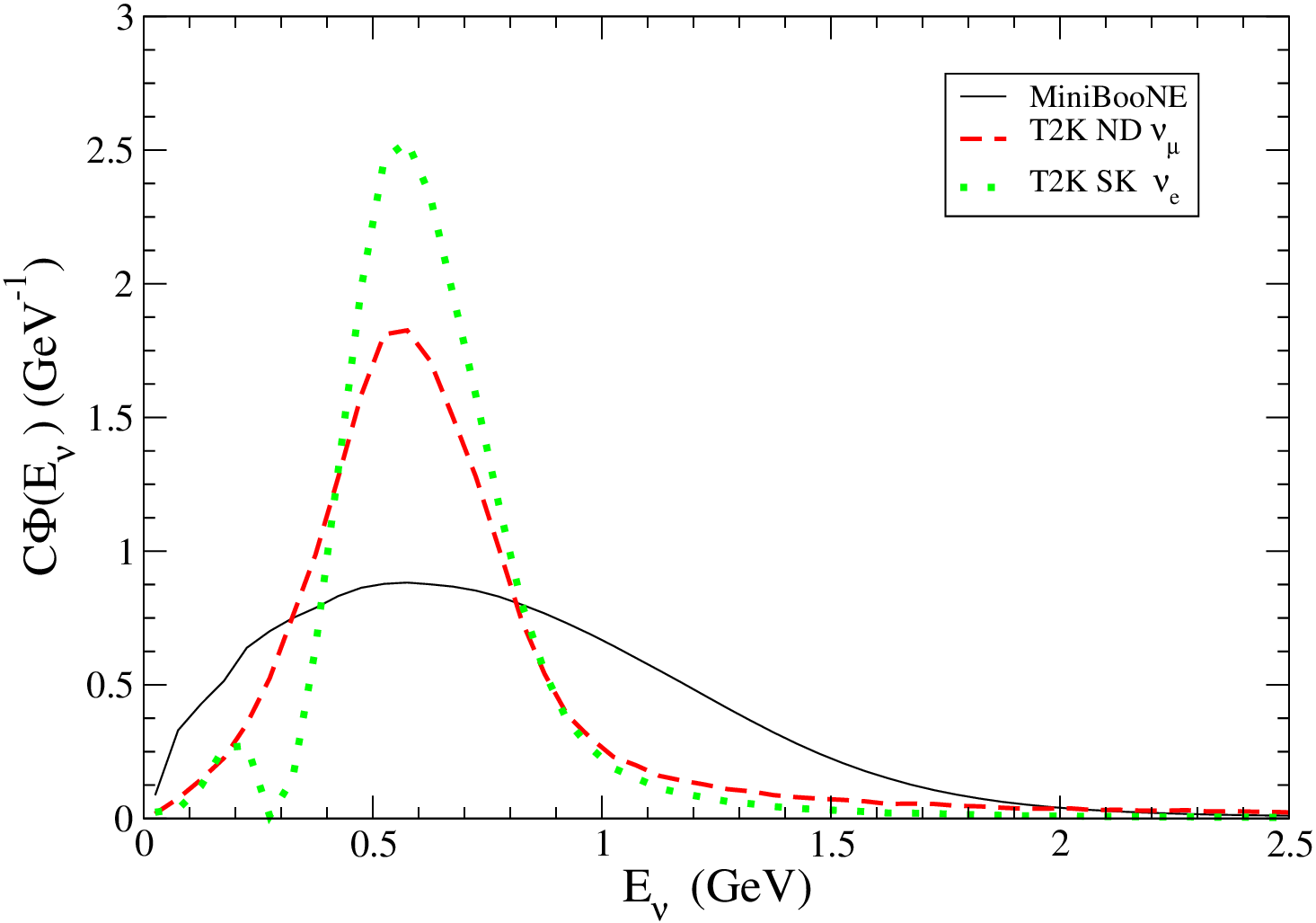}
\caption{(Color online) MiniBooNE and T2K normalized fluxes. The $\nu_{\mu}$ MiniBooNE flux is taken from Ref.\cite{AguilarArevalo:2010zc}. 
The $\nu_{\mu}$ T2K flux at near detector is taken from \cite{Abe:2012gx}. The $\nu_{e}$ T2K flux at Super Kamiokande far detector is obtained from Eq. (\ref{eq_t2k_sk_flux}).
}
\label{fig_minib_t2k_flux}
\end{center}
\end{figure}

Our work is based on the description of the neutrino cross section through the nuclear responses as is described in the work of Ref. \cite{Martini:2011wp} 
which successfully accounted for the MiniBooNE data on the double differential cross section \cite{AguilarArevalo:2010zc}. 
The problem of finding the probability  distribution of the neutrino genuine energy in the ``quasielastic'' events cannot be dissociated 
from the neutrino energy distribution which is an input in this evaluation, as each energy is weighted with the neutrino flux, 
$\Phi(E_{\nu})$. 
In the applications of the present work we take three examples of the neutrino energy distribution, related to the corresponding neutrino oscillations experiments. 
One is for the MiniBooNE circumstances \cite{AguilarArevalo:2007it,AguilarArevalo:2008rc}, 
i.e. for this neutrino flux energy distribution that we take from Ref. \cite{AguilarArevalo:2010zc}. 
For the second case  we consider the T2K flux \cite{Abe:2012gx} at the near detector (ND), $\Phi_{T2K}^{ND}(\nu_{\mu})$, 
which is somewhat more narrow. The energy profiles of the two beams are shown 
in Fig. \ref{fig_minib_t2k_flux}. Finally in the application of our results to the energy distribution of the T2K electron events as observed at the Super Kamiokande far 
detector \cite{Abe:2011sj}, 
we take the electron neutrino flux $\Phi_{T2K}^{SK}(\nu_e)$ from an evaluation in the oscillation description 
 \begin{equation}
\label{eq_t2k_sk_flux}
\Phi_{T2K}^{SK}(\nu_e)\propto \sin^2\left(\frac{\Delta m_{32}^2 L}{4 E_\nu}\right)\Phi_{T2K}^{ND}(\nu_{\mu}),
\end{equation}
where $|\Delta m_{32}^2| = 2.4 \times 10^{-3}$ eV$^2$ and $L=295$ km. This electron neutrino flux is also shown in Fig. \ref{fig_minib_t2k_flux}; it is definitely narrower that the MiniBooNE one or the T2K near detector one.
  
\section{Analysis and results}

\subsection{With specification of the lepton observables}
\begin{figure}
\begin{center}
  \includegraphics[width=16cm,height=12cm]{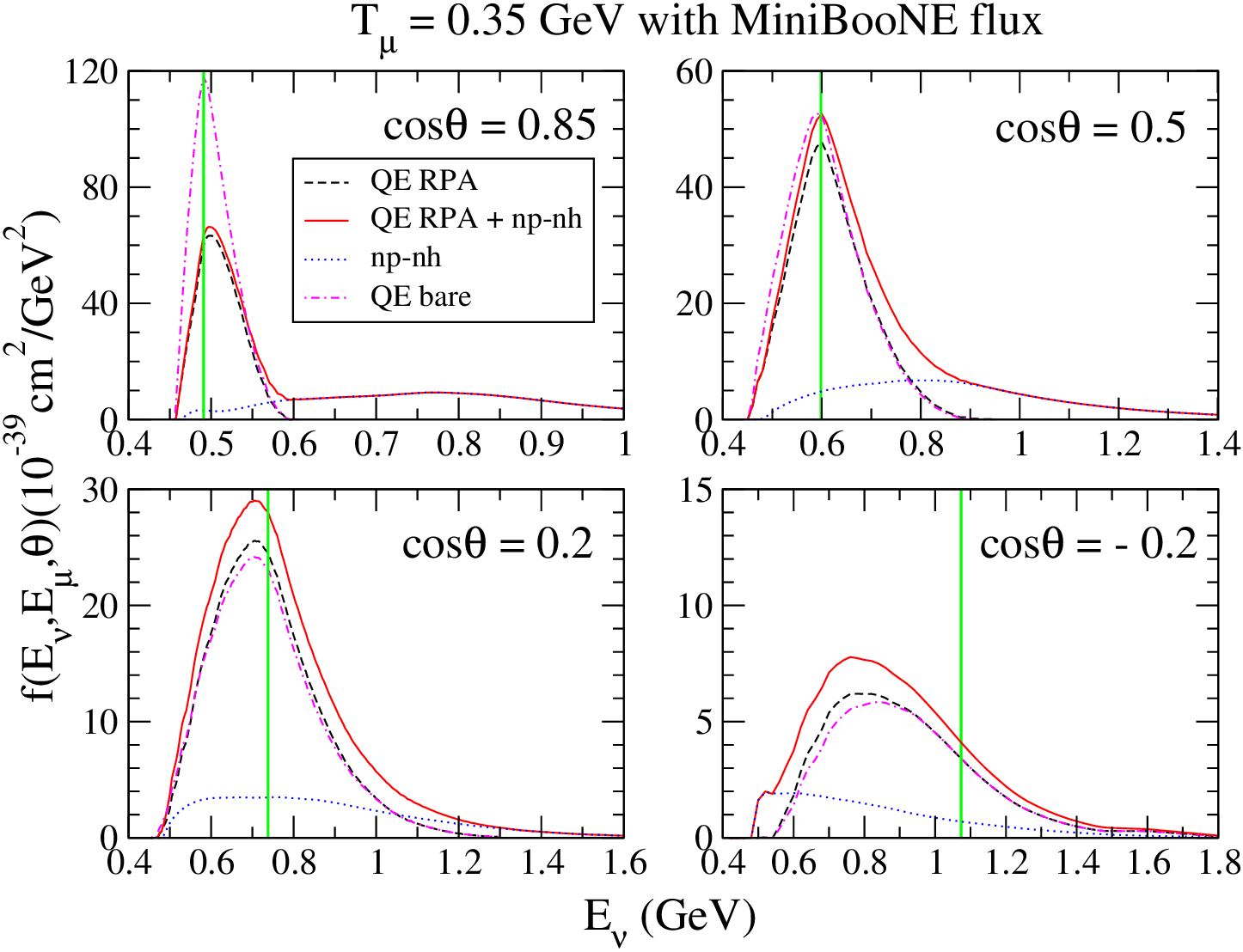}
\caption{(Color online) Probability distributions for $T_\mu$=0.35 GeV evaluated at various values of the emission angle for the MiniBooNE flux. 
The vertical lines give the position of the corresponding reconstructed energy $\overline{E_\nu}$ according to Eq. (\ref{eq1_erec}).} 
\label{fig_minib_d2s_phi_035}
\end{center}
\end{figure}
\begin{figure}
\begin{center}
  \includegraphics[width=16cm,height=12cm]{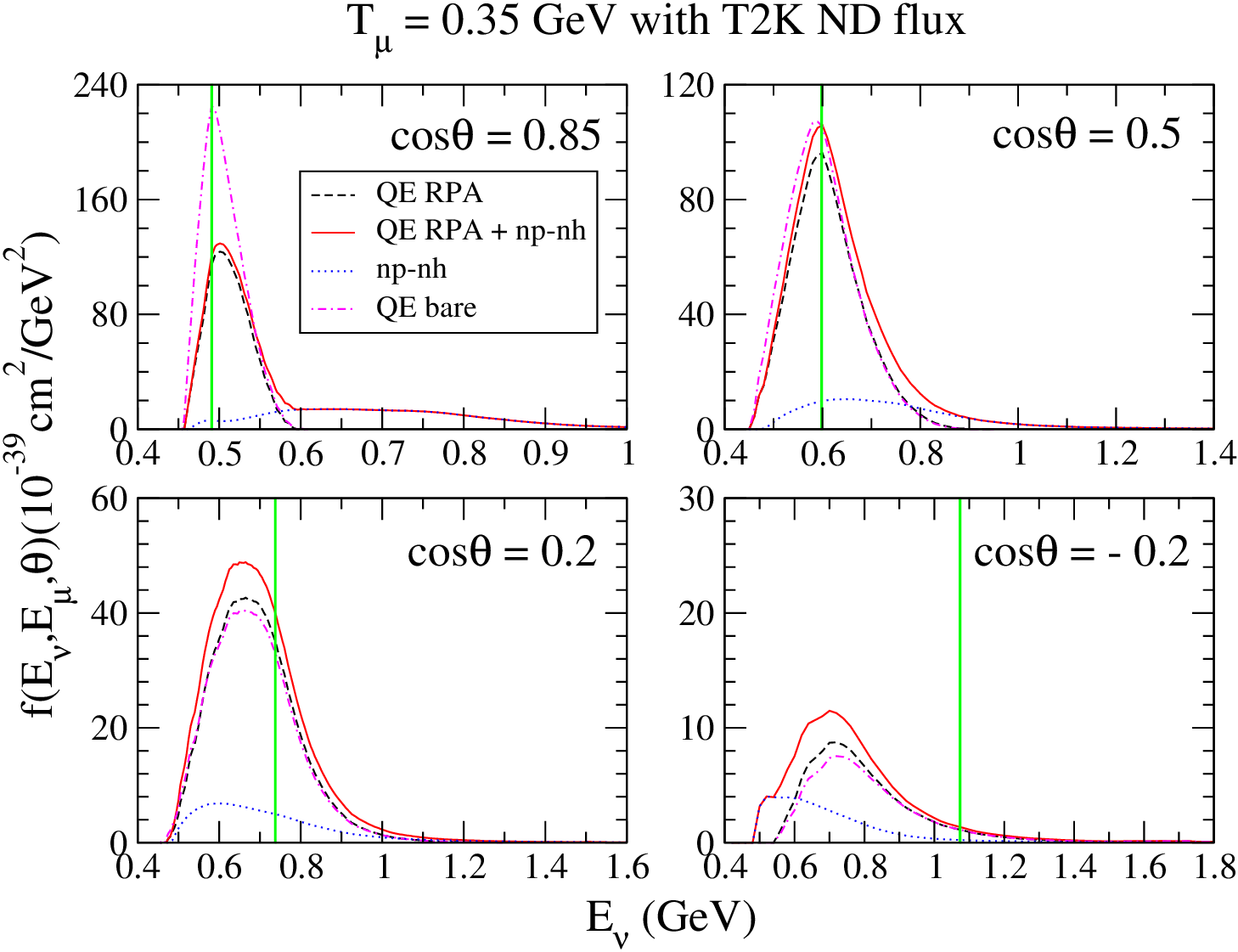}
\caption{(Color online) Probability distributions for $T_\mu$=0.35 GeV evaluated at various values of the emission angle for the T2K near detector flux. 
The vertical lines give the position of the corresponding reconstructed energy $\overline{E_\nu}$ according to Eq. (\ref{eq1_erec}).} 
\label{fig_t2k_d2s_phi_035}
\end{center}
\end{figure}

\begin{figure}
\begin{center}
  \includegraphics[width=16cm,height=12cm]{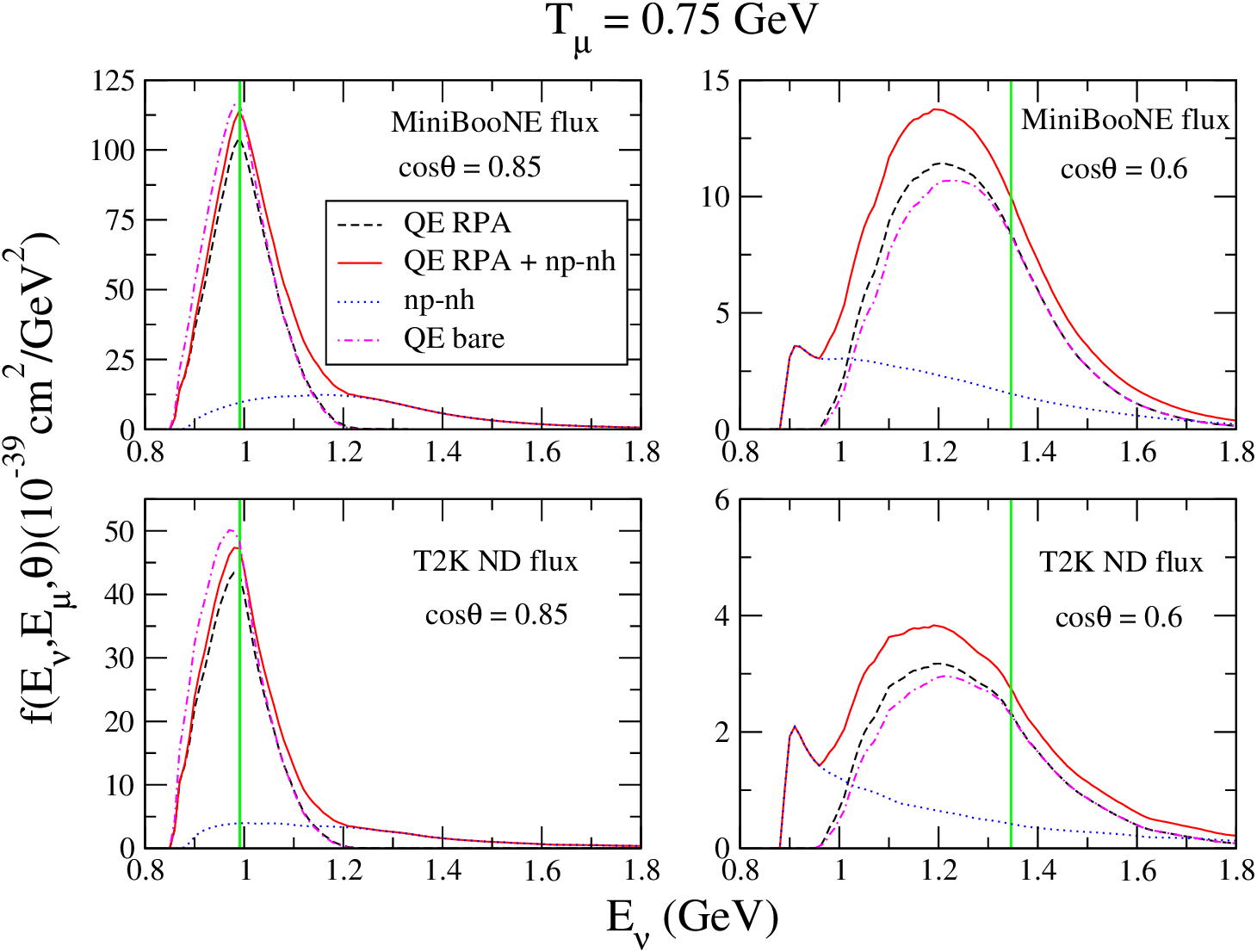}
\caption{(Color online) Probability distributions for $T_\mu$=0.75 GeV evaluated at various values of the emission angle for the MiniBooNE and T2K near detector fluxes. 
The vertical lines give the position of the corresponding reconstructed energy $\overline{E_\nu}$ according to Eq. (\ref{eq1_erec}).} 
\label{fig_minib_d2s_phi_075}
\end{center}
\end{figure}

We  first discuss  what is the neutrino energy distribution $f(E_{\nu}, E_{\mu }, \theta)$ for a given set of muon energy $E_{\mu }$ and angle $\theta$. In this case one explores a single hyperbola  in the $(\omega, q)$ plane. The neutrino energy distribution follows that of the energy $\omega$ transferred to the nuclear system along this hyperbola, as $ E_\nu = E_\mu+\omega$, with a weight proportional to the neutrino flux at this energy  $\Phi( E_{\nu })$.  Thus it is defined as
 \begin{equation}
 \label{f_E_E_theta} 
f( E_{\nu }, E_{\mu }, \theta) d E_{\nu }= C
\left[\frac{d^2 \sigma}{d \omega  ~d\mathrm{cos}\theta}\right]_{\omega=E_{\nu}-E_{\mu}} \Phi(E_{\nu}) d E_{\nu },
\end{equation}
 where $C$ is a normalization constant. 
 Several choices are possible for the normalization and we take here for $C$ the inverse value of the total flux, 
 $C= ( \int d E_{\nu} \Phi(E_{\nu}))^{-1}$, in such a way that the probability $f$ has the dimensions of a cross section divided by an energy squared.
 The neutrino cross section which enters Eq. (\ref{f_E_E_theta}) has two pieces, the one particle-one hole (1p-1h) part, which is the genuine quasielastic part, and the multinucleon emission (np-nh) 
one which represents a sizeable fraction of the first one. 
The second part in particular introduces a high energy tail in the neutrino cross section, ignored in the usual reconstruction procedure which assumes 
that all of the cross section for this  energy and angle is concentrated at the intercept of the corresponding hyperbola with the quasielastic line. For all numerical evaluations of this article we use the description of the nuclear responses of our previous work \cite{Martini:2011wp}. 
As in this work, relativistic corrections to the nuclear responses are incorporated, the 1p-1h component of the response 
is also treated in the random phase approximation (RPA). 
The multinucleon component is the same as the one used in Ref. \cite{Martini:2011wp}, which is described in Refs. 
\cite{Martini:2009uj,Martini:2010ex}.

In our evaluation we do not introduce any experimental cut-off which may be necessary to describe actual data. 
In this work we consider charged current neutrino cross sections on carbon which is the element constituting the MiniBooNE detector and the T2K near detector. 
In the case of Super Kamiokande the element involved is the oxygen but in Ref. \cite{Martini:2009uj} we have shown that the cross sections per nucleon 
calculated for carbon and oxygen are almost identical both in the quasielastic and multinucleon channel. As a consequence we perform the calculations for carbon.
   
The outcome of our study is illustrated in 
Figs. \ref{fig_minib_d2s_phi_035} and \ref{fig_t2k_d2s_phi_035} which represent the probability distributions for one value of the muon energy, $E_\mu=0.45$ GeV  
($T_\mu= 0.35$ GeV) and evaluated at various values of the emission angle.
Figure \ref{fig_minib_d2s_phi_075} represents the same quantities for $T_\mu= 0.75$ GeV.   The corresponding hyperbolas for this particular muon energy and the selected values of the muon angle are displayed in Fig. \ref{fig_hyperbolas}. We have performed the evaluations  for the two neutrino fluxes, MiniBooNE and T2K at the near detector. 
The general features of the two distributions are rather similar in  the two cases, with some differences that we will comment. 
They should be compared to the distributions for a dilute Fermi gas which is a delta function at the value of the reconstructed energy,  
$\overline{E_\nu}$,  indicated by a line in  Figs. \ref{fig_minib_d2s_phi_035}, \ref{fig_t2k_d2s_phi_035} and \ref{fig_minib_d2s_phi_075}. 
The partial probabilities for 1p-1h and np-nh are shown separately, as well as the total one. 
The features of the probability distribution are radically different according to the value of the emission angle. 
The genuine quasielastic part, i.e., the 1p-1h part  itself, can display sizeable deviations from the reconstructed energy. 
At small angles the corresponding probability distribution is not symmetrical around $\overline{E_\nu}$, and it is hardened. 
Part is due to the Pauli blocking effect which cuts the responses at small momenta. The RPA also has some hardening and quenching effect at small angles as 
it is displayed in Figs. \ref{fig_minib_d2s_phi_035}, \ref{fig_t2k_d2s_phi_035} and \ref{fig_minib_d2s_phi_075} 
where the probabilities with and without RPA are shown. Moreover the np-nh part creates a high energy tail in the nuclear response, 
above the quasielastic peak, quite visible in the transverse part of the inclusive electron scattering data \cite{Alberico:1983zg} \footnote[1]{A comparison between the electron scattering data on carbon and our model can be found in Ref. \cite{Martini:2011ui}. The agreement is satisfactory.}.
Larger energy transfer, $\omega$, meaning larger neutrino energies, the distribution acquires a high-energy tail, absent for an uncorrelated Fermi gas, 
which produces an appreciable increase of the mean neutrino energy. 
With the increasing muon angle the hardening of the distribution disappears at first and the 1p-1h distribution is centered near the reconstructed energy value. 
The np-nh component still creates the high energy tail.

For larger angles an unexpected feature emerges, the energy distribution becomes softened with respect to the reconstructed energy. In the backward directions its distribution seems even unrelated to the reconstructed energy value. This feature can be understood from the corresponding hyperbola of Fig. \ref{fig_hyperbolas}. 
With the increasing muon angle the portion of the hyperbola inside the quasielastic region becomes very large. For instance for $T_{\mu}=0.35$ GeV 
already at a value of $\cos\theta= 0.2$ the portion lies between the momentum transfer $q\approx 0.6$ GeV and  $q\approx 1$ GeV. 
In such a situation the form factors, vector or axial, have a strong suppression effect at the largest momenta (or energies), 
which explains the distortion of the neutrino energy distribution towards smaller energies which can be in addition  favored by the flux factor. 
The problem becomes even more pronounced in the backward directions. For instance for $\cos\theta= -0.2$ the distribution is concentrated at 
low energies and has little to do with the reconstructed energy, in particular for T2K (Fig. \ref{fig_t2k_d2s_phi_035}). 
These large angles correspond to very large values of the momentum transfers, for which the np-nh component, which is difficult to evaluate reliably at these momenta, 
is not the main cause of the distortion effect. Notice that all the features mentioned above can be enhanced or diminished depending on the region of energy of the neutrino spectrum, which may or not favor the larger energies such as the np-nh tail. At small  $E_\mu$ values, such as 0.45 GeV, the tail for a value $\cos\theta=0.85$ occurs in an energy region of increasing  flux for MiniBooNE and is favored while this is not so for T2K, which explains the difference in the importance of the tail.

\subsection{Distributions with no specification of the lepton observables}
\begin{figure}
\begin{center}
  \includegraphics[width=12cm,height=8cm]{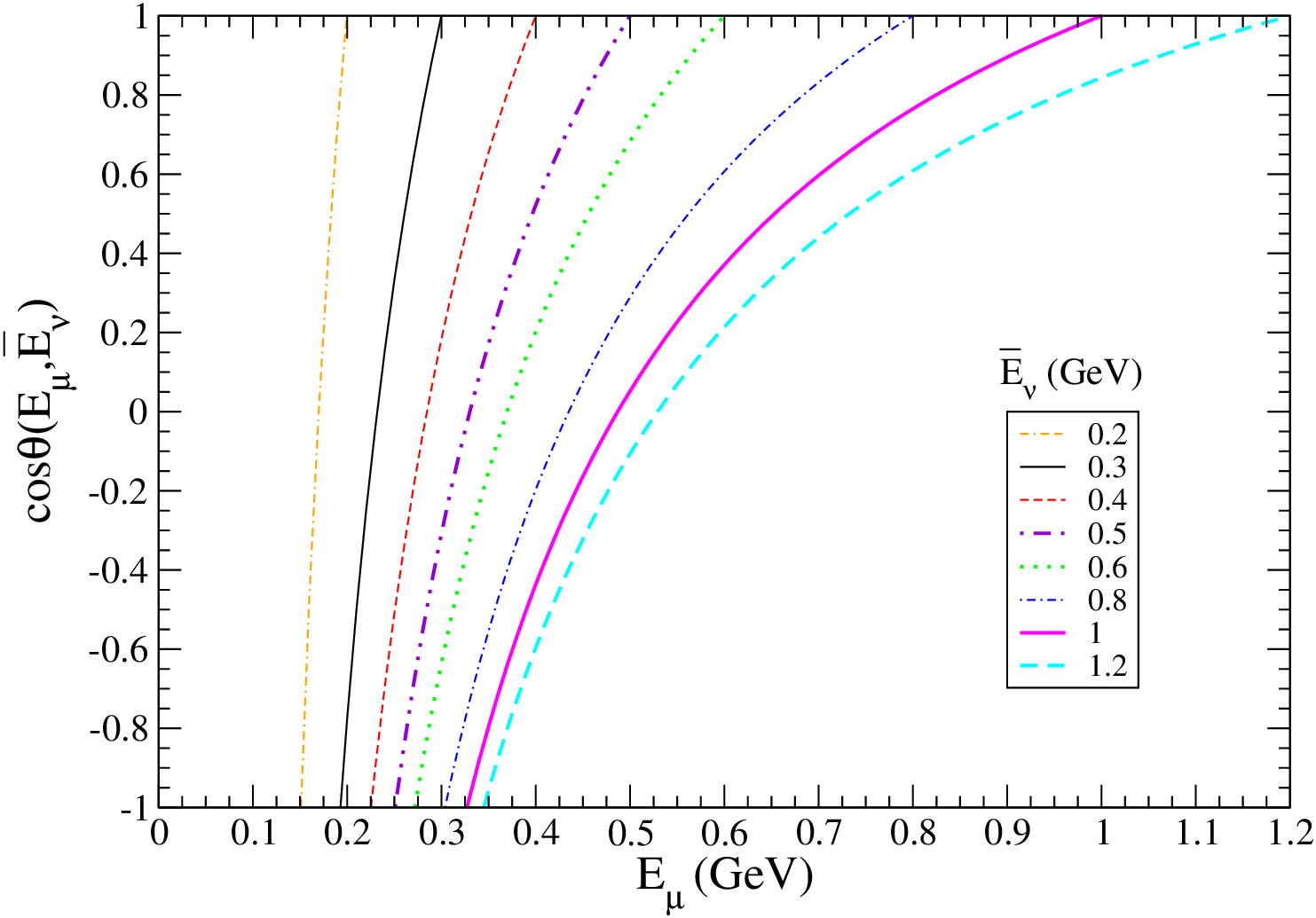}
\caption{(Color online) Cosine of the muon angle solution of the Eq. (\ref{eq4_ebar}) 
for various values of the reconstructed neutrino energy as a function of the muon energy.}
\label{fig_cos_rec_emu}
\end{center}
\end{figure}

In the previous study we have derived the  probability distribution $f( E_{\nu}, E_{\mu},\theta)$ for  events with fixed muon variables: energy and emission angle. 
In the following we investigate a more global quantity, namely the neutrino  energy distribution  for a given neutrino reconstructed energy,  $F( E_{\nu }, \overline{E_{\nu})}$, without any explicit reference to the muon variables. Many couples of values of these variables, muon energy and emission angle, can lead to the same reconstructed energy and one has to sum over these couples. A given value of the reconstructed energy, $\overline{E_\nu}$ links the muon angle and its energy through
\begin {equation}
\label{eq4_ebar}
 \overline{E_{\nu}} P_{\mu} \cos\theta+M (\overline{E_{\nu}}-E_{\mu})-\overline{E_{\nu}}E_{\mu }+m^2_{\mu}/2 =0.
\end {equation} 
We denote by $\cos\theta(E_{\mu} , \overline{E_{\nu}})$ the cosine of the muon angle solution of this equation for a given set of values, $E_\mu$ and  $\overline{E_\nu}$. 
Figure \ref{fig_cos_rec_emu} displays its evolution with the muon energy for various values of the reconstructed energy. 
The probability energy distribution, $F(E_{\nu},\overline{E_{\nu}})$,  is  obtained as an integral  over the muon energy of the double differential neutrino cross section, taken at  this  particular value of the muon angle

\begin{equation} 
F(E_\nu, \overline{E_\nu}) = c\frac{ \Phi(E_{\nu})}{  \int d E_{\nu} \Phi(E_{\nu})}
 \int _{E_\mu^{\textrm{min}}} ^{E_{\mu}^{\textrm{max}}}~d E_{\mu}
\left[\frac{d^2 \sigma}{d \omega  ~d\mathrm{cos}\theta}\right]_{\omega=E_{\nu}-E_{\mu},~\mathrm{cos}\theta=\mathrm{cos}\theta(E_\mu, \overline{E_\nu})} = c I(E_\nu, \overline{E_\nu}),
\end{equation}
where the second identity  defines the quantity $I$, $E_\mu^{\textrm{min}}(^{\textrm{max}})$ are the extreme values of $E_{\mu}$ obtained from Eq. (\ref{eq4_ebar}) 
by making $\cos\theta=\pm1$
and $c$ is a normalization constant that we choose in order to have a total probability unity,
$ \int d E_{\nu}~F( E_\nu,\overline{E_\nu}) = 1$, which gives to $F$ the dimensions of an inverse energy.

\begin{figure}
\begin{center}
  \includegraphics[width=16cm,height=12cm]{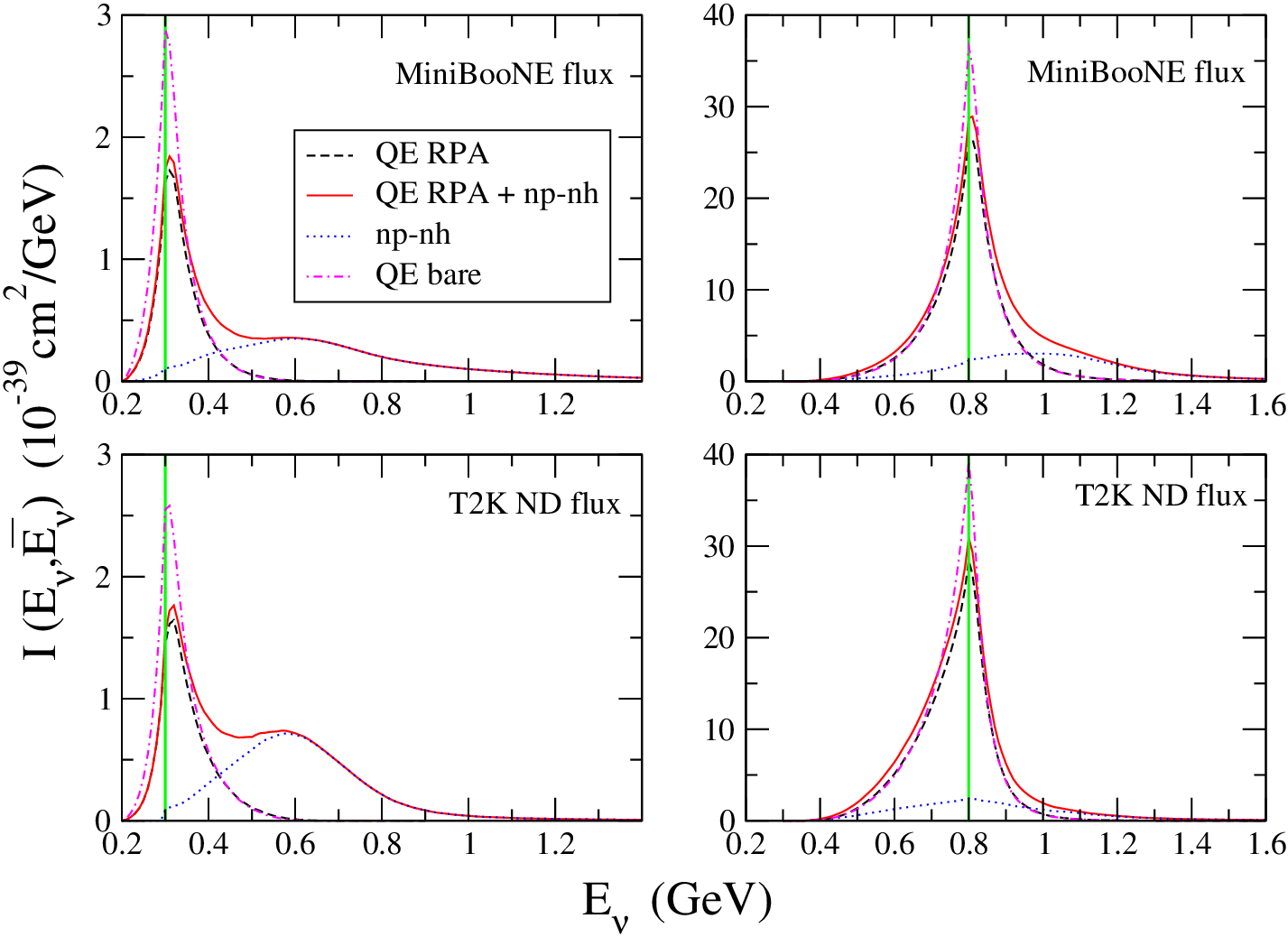}
\caption{(Color online) Probability distribution before normalization, $I$, for  several $\overline{E_\nu}$ values corresponding to the different vertical lines. 
The MiniBooNE and T2K near detector fluxes are used.}
\label{fig_ferec_enu_300_1200_minib.eps}
\end{center}
\end{figure}

\begin{figure}
\begin{center}
  \includegraphics[width=16cm,height=12cm]{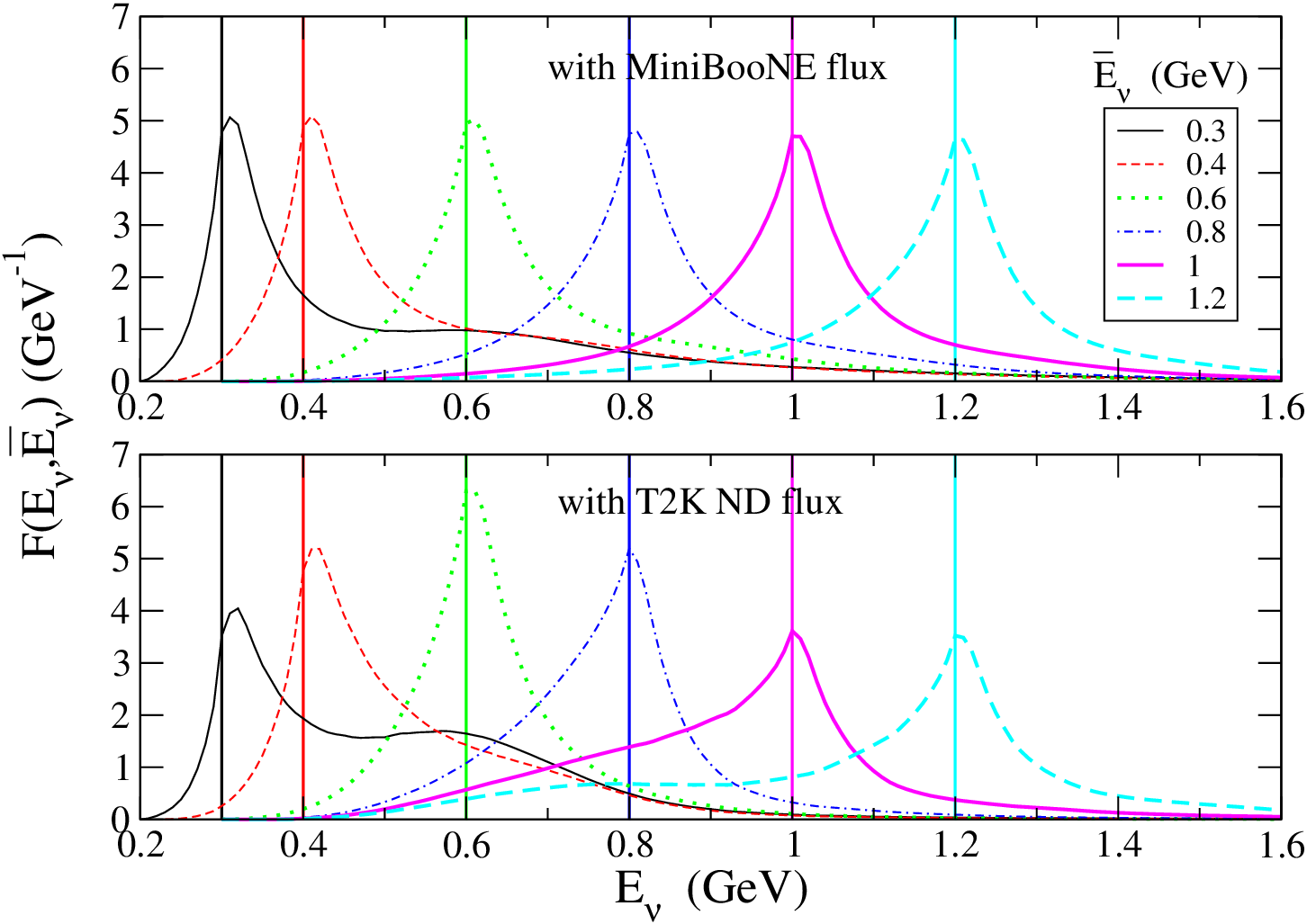}
\caption{(Color online) Probability distributions for several $\overline{E_\nu}$ values corresponding to the different vertical lines. 
Upper panel: using the MiniBooNE flux. Lower panel: using the T2K near detector flux.}
\label{fig_norm_minib_t2k}
\end{center}
\end{figure}

\begin{figure}
\begin{center}
  \includegraphics[width=16cm,height=6cm]{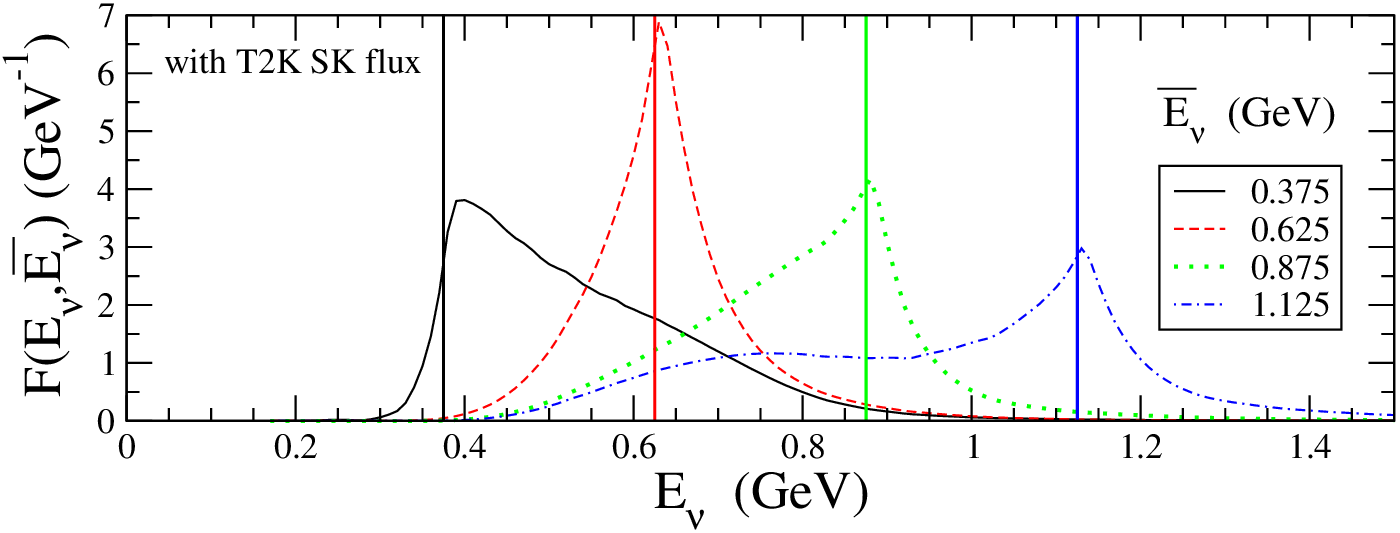}
\caption{(Color online) Probability distributions for several $\overline{E_\nu}$ values, corresponding to the different vertical lines, using the T2K flux 
predicted at Super Kamiokande for electron neutrinos.}
\label{fig_norm_t2k_far}
\end{center}
\end{figure}

\begin{figure}
\begin{center}
  \includegraphics[width=16cm,height=12cm]{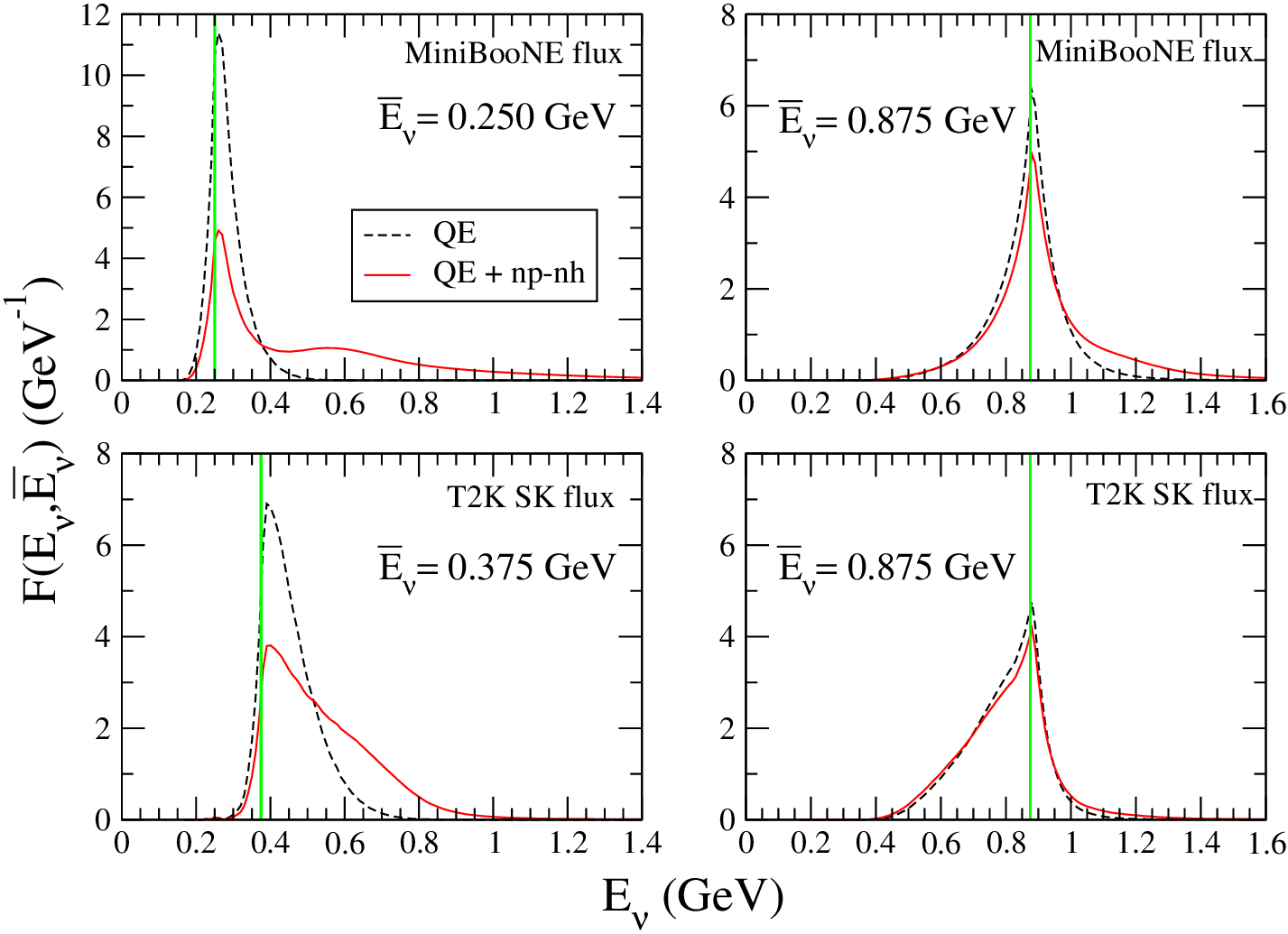}
\caption{(Color online) Normalized probability distribution for some $\overline{E_\nu}$ values corresponding to the different vertical lines. 
The MiniBooNE flux and the T2K far detector Super Kamiokande $\nu_e$ flux are used.}
\label{fig_F_norm_minib_t2k}
\end{center}
\end{figure}

Some examples of these distributions are shown in Fig. \ref{fig_ferec_enu_300_1200_minib.eps} for some $\overline{E_\nu}$ values. 
Here we have plotted the quantity $I(E_\nu, \overline{E_\nu})$ without the normalization factor $c$, instead of the probability density $F$,  
in order to single out the various contributions to the cross sections, quasielastic with and without RPA, and the multinucleon contribution.
The main features are similar to those found previously for fixed muon variables. This distribution has a high energy tail due to the np-nh cross section. For small or moderate  $\overline{E_\nu}$ values the distribution 
is not symmetrical around  $\overline{E_\nu}$. The 1p-1h part itself is shifted towards larger neutrino energies by the  Pauli blocking effects and the RPA influence. 
In addition the np-nh component is responsible for the high energy tail. For the smallest values of the reconstructed energy, 
$\overline{E_\nu} \lesssim 0.5$ GeV, the np-nh tail is  favored by the energy evolution of the neutrino flux which increases in this energy region, 
in particular for the T2K energy profile. 
With increasing values of the reconstructed energy the RPA hardening of the 1p-1h disappears and the form factor effect softens this part as explained previously. 
The np-nh compensates in part for this softening, and the overall distribution looks approximately symmetric in particular with the MiniBooNE flux. For larger 
$\overline{E_\nu}$ values the distribution becomes again asymmetrical but it is now the low energy side which is favored,  largely due to the flux effect.
These features are apparent in Fig. \ref{fig_norm_minib_t2k} where several probability distributions, $F(E_\nu, \overline{E_\nu})$, now normalized, 
are plotted in the same figure, showing the progressive evolutions with increasing values of the reconstructed energy. As illustrated in Fig. \ref{fig_norm_t2k_far}, this asymmetry effect is even more pronounced with the T2K far detector electron neutrino flux which is very narrow, as was shown in Fig. \ref{fig_minib_t2k_flux}.
The asymmetry of these distributions which favors higher energies at low $\overline{E_\nu}$ values  and  smaller energies
for large $\overline{E_\nu}$  values has  a narrowing  effect for the smeared distribution as compared to the unsmeared one,
as will be illustrated in the application to the T2K data. For the MiniBooNE case, with the assumption of a broad flux, the main effect is the transfer of low reconstructed energy events towards larger energies. This reshaping arises from the np-nh tail of the distribution. This is confirmed by the comparison of the normalized probability distributions with and without the multinucleon contribution as shown in Fig. \ref{fig_F_norm_minib_t2k}. 
For small values of the reconstructed energy the distribution in $E_\nu$ which incorporates the np-nh contribution is sizably deformed with respect to the one where this contribution is not included, it is suppressed at low  $E_\nu$ values and enhanced at large ones. This hardening effect is due to the important role of the np-nh tail in the normalization factor. It has interesting consequences for the analysis of the experimental data, as will be discussed in the next section.  

  The average  neutrino energy, which depends on the reconstructed energy, $(E_{\nu})_{\textrm{average}}(\overline{E_\nu})$, 
is related to the normalized probability distribution $F$ by
\begin{equation}
\label{eq_en_average}
(E_{\nu})_{\textrm{average}}(\overline{E_\nu}) =  \int d E_{\nu}~E_{\nu}~F(E_\nu, \overline{E_\nu}).
\end{equation} 
The comparison with the reconstructed value  gives an idea of the validity of the narrow quasielastic approximation. However in view of the shape of the distribution with the high-energy tail, it would be meaningless to replace in the analysis of experimental data the reconstructed energy by the average one, only the energy distribution is significant. In the Table \ref{table} some examples are given for the average neutrino energy of Eq. (\ref{eq_en_average})
for various values of  $ \overline{E_\nu} $ both for the MiniBooNE and the T2K near detector  neutrino energy profiles. We also give the same quantity but without np-nh contribution. 
Even in this case there is some slight modification at low $\overline{E_\nu}$ values: the average neutrino energy is increased due to the RPA hardening of the responses.
\begin{table}[t]
    \begin{center}
        \begin{tabular}{c|cc|cc}
            \hline
                                  & MiniBooNE &  &  T2K ND  &   \\
   $\overline{E_\nu}$ (MeV)       & QE + np-nh  & QE  & QE + np-nh & QE\\
            \hline
            300 &546& 335  &514 & 350  \\
            400 &579& 435  &529 & 446  \\ 
	    500 &638& 527  &575 & 528  \\ 
	    600 &711& 619  &638 & 606  \\ 
	    800 &861& 799  &781 & 758  \\ 
	    1000&1024& 981 &937 & 914  \\ 
            1200&1190& 1164&1116& 1104 \\ 
            \hline
        \end{tabular}
    \caption{    
    \label{table} The average neutrino energy, Eq. (\ref{eq_en_average}), in MeV for various values of  $ \overline{E_\nu} $ 
considering the MiniBooNE and T2K near detector neutrino fluxes. The results are obtained 
with (QE+np-nh columns) and without (QE columns) the multinucleon contribution in the cross sections.}
    \end{center}
\end{table}
\section{Applications to actual data on electron neutrino appearance}
Previous formulas given for muon emission apply for electron neutrinos, replacing $m_\mu$ by $m_e$. 
The results are in fact quite similar. For  given lepton energy and angle the reconstructed energy is practically identical. 
For the same neutrino energy profiles the neutrino energy probability distributions are also quite similar in the two cases. 
In the following we apply our work to actual neutrino data of MiniBooNE \cite{AguilarArevalo:2008rc} and T2K \cite{Abe:2011sj} on the observation of an electron excess in 
a beam of muon neutrinos.  
The experimental collaborations give  the distribution of electron excess events in  terms of the electron neutrino reconstructed energy. These neutrinos are attributed 
to the oscillation of muon neutrinos. 
As the oscillation phenomena depends on the neutrino energy it is important to have the distribution in terms of the real neutrino energy and not the reconstructed one which is the aim of this work. In a forthcoming work we will discuss the angular distribution problem and the $Q^2$ distribution. 
In our approach the transformation which leads to the true neutrino energy distribution is readily performed applying our previous probability distribution $F$. The number of events $g(\overline{E_\nu})d \overline{E_\nu}$ in a bin of reconstructed energy $ \overline{E_\nu}$ should be smeared by the function 
$F(E_\nu, \overline{E_\nu})$ leading to the distribution $G(E_\nu)$ in terms of the real neutrino energy
\begin{equation} 
\label{eq_G}
G(E_\nu)=\int d  \overline{E_\nu} g(\overline{E_\nu}) F(E_\nu,\overline{E_\nu}).
\end{equation} 
 As the function $F$ is normalized to unity this transformation conserves the area:
\begin{equation} 
 \int d  \overline{E_\nu} g(\overline{E_\nu})= \int {d  {E_\nu} G(E_\nu}).
\end{equation} 

 We have performed this transformation both for the MiniBooNE and the T2K Super Kamiokande distributions.
The evaluation of the smearing function $F$ involves the electron neutrino energy spectrum. In the T2K case we have given the evaluation of this quantity from the muon neutrinos spectrum  and the oscillation formula of Eq. (\ref{eq_t2k_sk_flux}). For MiniBooNE we make for the moment the assumption that the energy distribution is the same as the initial muon one. In the future we will improve this approximation in an oscillation model.
As a last remark notice that our procedure is reversible: starting from a  neutrino distribution in terms of the true energy, for instance a theoretical one 
as will be given for T2K, we could apply a similar procedure to derive the corresponding distribution in terms of the reconstructed energy, directly comparable to the experimental data.

\subsection{MiniBooNE}
\begin{figure}
\begin{center}
  \includegraphics[width=12cm,height=8cm]{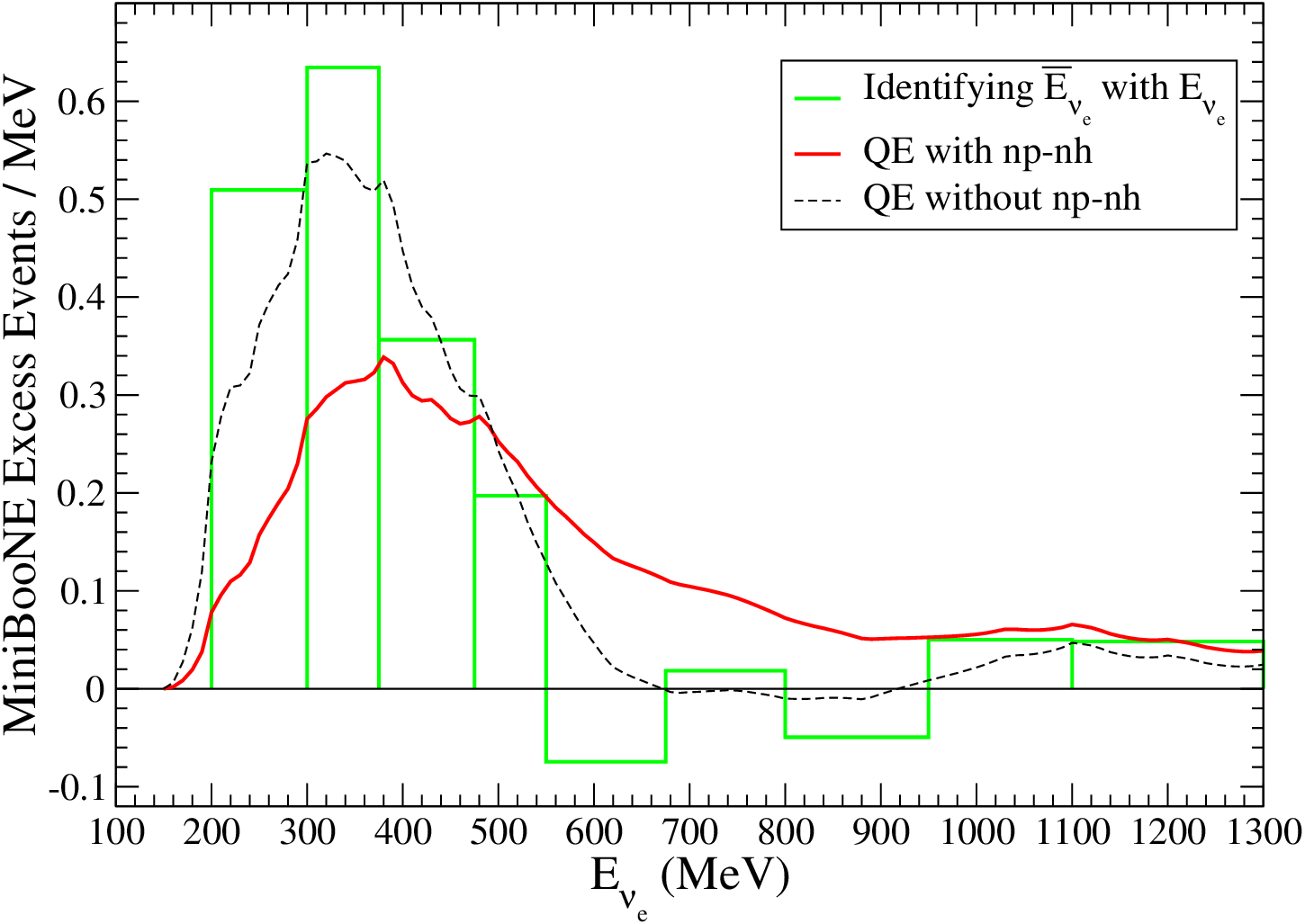}
\caption{(Color online) Three different treatments of experimental MiniBooNE data \cite{AguilarArevalo:2008rc} on electron neutrino excess events. 
The histogram is obtained identifying the reconstructed neutrino energy $\overline{E_\nu}$ with the true neutrino energy $E_\nu$. 
The full line is our smeared distribution including the multinucleon contribution. 
The dashed line is our smeared distribution without the multinucleon contribution.}
\label{fig_minib_excess}
\end{center}
\end{figure}
We first start with the MiniBooNE case where the so-called ``low energy MiniBooNE anomaly'' has been displayed \cite{AguilarArevalo:2007it,AguilarArevalo:2008rc} 
with a number of electron event excess 
concentrated in the low-energy region of the reconstructed energies. The MiniBooNE Collaboration \cite{AguilarArevalo:2007it} 
has  proposed  to discard the low-energy points responsible for the anomaly, which do not fit easily the oscillation interpretation. 
But there is no obvious reason to do so and here we keep all the points which possibly  represent an electron excess. 
The experimental results are given in terms of an excess number in large intervals of  $\overline{E_\nu}$ (200-300MeV, 300-375 MeV, etc). 
In view of this and for simplification, instead of the continuous integral on  $\overline{E_\nu}$ of Eq. (\ref{eq_G}),  
we have taken for $F$ a discrete sum of three  energy points in each bin, the extreme and the central ones (which explains the spikes of our curves). 
The result of our study is contained in Fig. \ref{fig_minib_excess}. The full curve represents our smeared distribution obtained from the experimental data excess 
\cite{AguilarArevalo:2008rc}. The experimental histogram is also shown in Fig. \ref{fig_minib_excess} 
under the assumption $\overline{E_\nu}\equiv E_\nu$. 
 For information we also give the smeared distribution in the absence 
of the np-nh contribution (dashed curve) which is more similar to the original data. However for us the more realistic curve is the one with the inclusion of the np-nh component. We stress that all the distributions shown in  Fig. \ref{fig_minib_excess} represent the same experimental data with the different 
treatment of the abscissa, the neutrino energy. The areas under these curves are the same. The salient feature is  that our smeared distribution (with the np-nh component) 
is broadened and hardened, with the suppression of the low energy excess, as compared to the initial histogram. 
The effect arises from  the high energy tail of our smearing function. The difference with the original distribution is so large that it could affect the interpretation in terms of oscillation. Our smeared distribution with the np-nh component is closer to the LSND \cite{Aguilar:2001ty} solutions shown in Ref. \cite{AguilarArevalo:2008rc} than the original histogram. At present this important conclusion is only tentative as the flux evaluation should and will be improved.

\subsubsection{Discussion and tests}

In view of the possible implications of this work for the  neutrino oscillation interpretation, several questions have to be raised. We have pointed out the uncertainties of the present evaluation  of the smearing function linked to the choice of the electron neutrinos energy distribution. This  will be improved.  Another point is that in the 
MiniBooNE  experiment there is a large experimental error on the number of interesting events linked to the existence of a large background which has to be subtracted. This subtraction is the reason for the negative value of the distribution in some bins, that we have kept in our evaluation. This background  is also given as a function of the reconstructed energy  $\overline{E_\nu}$. What part of the background should be smeared in the same way and what would be the influence on the distribution of interest?  This should be for instance the case for the background due to electron neutrinos originally present in the beam. This question should be cleared.

On the theoretical side the main question is: how reliable is our  smearing ? The hardening effect arises from the high-energy tail, due to the multinucleon contribution, of the smearing function for low  $\overline{E_\nu}$ values. It may change somewhat with different energy distributions of the electron neutrinos but we believe that the suppression of the low-energy peak with a shift towards higher 
neutrino energies is a robust prediction that will survive the better description of this distribution. 
Independently of the flux problem how reliable is our evaluation of this part?   We recall that the basic input for the evaluation of the smearing functions is the double differential cross section,  $ \frac{d^2 \sigma}{d \omega  ~d\mathrm{cos}\theta}$. We have tested this quantity \cite{Martini:2011wp} in connection with 
the experimental MiniBooNE data \cite{AguilarArevalo:2010zc}  on the double differential cross section  with respect to the muon observables, 
where it also enters. Our theoretical evaluation of $ \frac{d^2 \sigma}{dE_{\mu}  ~d\mathrm{cos}\theta}$ 
has been quite successful in the reproduction of these data. The multinucleon component is an important part of this cross section and is needed to reproduce the data 
in all the kinematical domains \cite{Nieves:2011yp},\cite{Martini:2011wp}. 
Note that Nieves \textit{et al.} \cite{Nieves:2011yp}  account for the MiniBooNE double differential cross sections with smaller contributions from the 
multinucleon channel but with an overall renormalization of the data by about $10\%$. A different proportion of the np-nh contribution may change the smearing distribution quantitatively but the general features of the distributions are expected to survive. The double differential cross section  involves a different weighting than the distribution $F$, 
by the muon neutrino MiniBooNE energy distribution
 \begin{equation}
 \label{cross}
\frac{d^2 \sigma}{dE_{\mu}~d\mathrm{cos}\theta}=
\frac{1}
{ \int \Phi(E_{\nu})~d E_{\nu}}
 \int ~d E_{\nu}
\left[\frac{d^2 \sigma}{d \omega  ~d\mathrm{cos}\theta}\right]_{\omega=E_{\nu}-E_{\mu}} \Phi(E_{\nu}).
\end{equation}
Hence for a given value of $E_{\mu}$ several energy transfer are involved. Although the link with the present evaluation is  therefore not a direct one we can argue that the success of our previous description \cite{Martini:2011wp} can give confidence to the present evaluation, as explained in the following. 
The suppression of the low energy peak of the MiniBooNE distribution of  Fig. \ref{fig_minib_excess} comes from the behavior of the smearing function for low values of 
the reconstructed energy ($\overline{E_\nu}\simeq$ 0.2 to 0.4 GeV)  with its high energy tail 
(see the corresponding distributions of Figs. \ref{fig_norm_minib_t2k} and \ref{fig_F_norm_minib_t2k}).
  We thus consider the example of  the smearing function for the value of the reconstructed energy, $\overline{E_\nu}=0.3$ GeV, 
and a large value of the  real energy, $ E_{\nu}=$0.7 GeV, which is well in the tail region and hence dominated by the multinucleon component that we want to test. We investigate what is the relevant range of the energy transfer for these two values of the neutrino energies. 
\begin{figure}
\begin{center}
  \includegraphics[width=9cm,height=6cm]{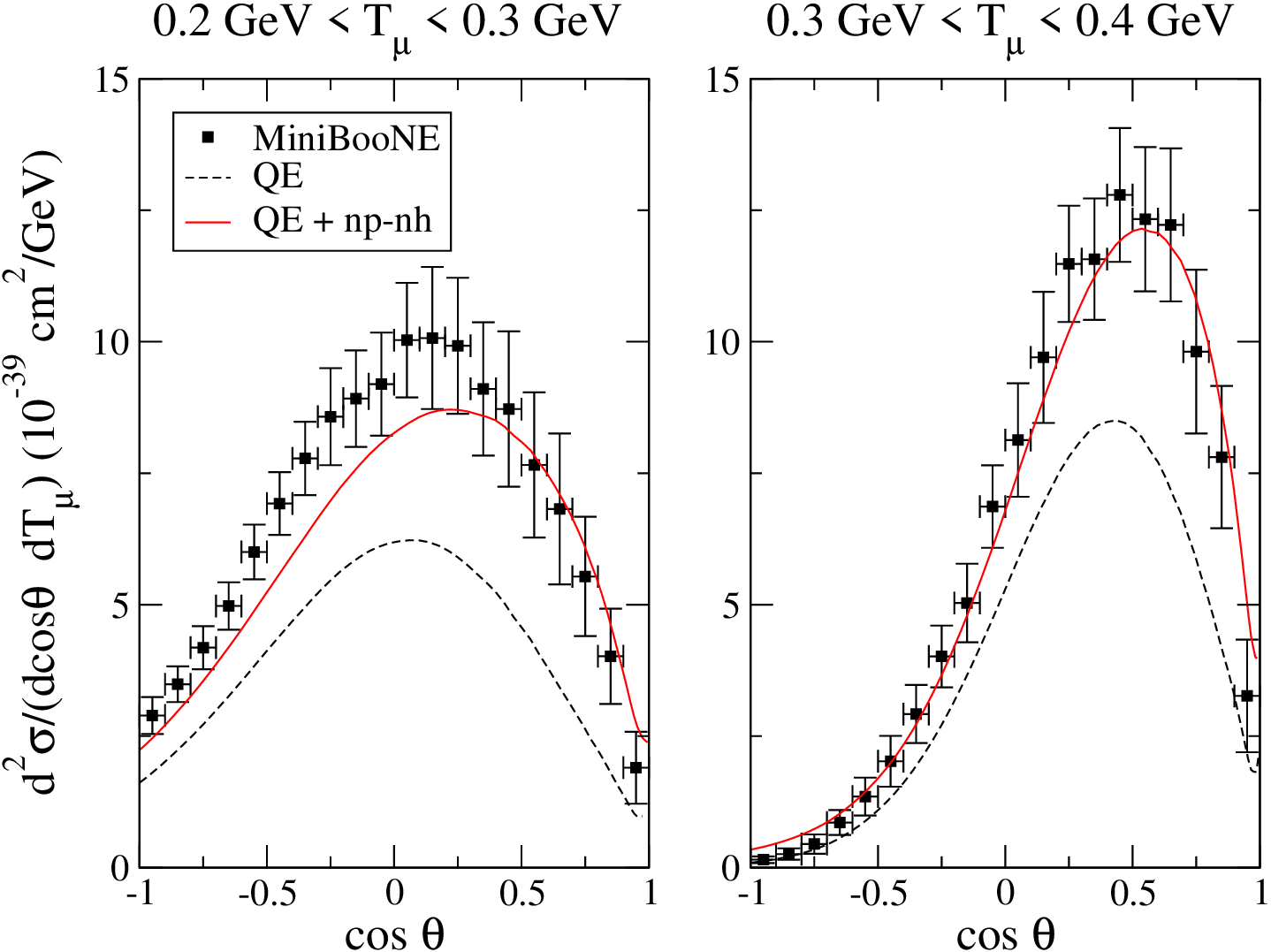}
\caption{(Color online) MiniBooNE flux-averaged charged current ``quasielastic'' $\nu_\mu$-$^{12}$C double differential cross section per neutron for some values of muon kinetic energy as a function 
of the scattering angle. The theoretical pure quasielastic (dashed curve) and the (solid curve) 
generalized quasielastic (QE+np-nh) cross sections are those obtained in \cite{Martini:2011wp}. The  MiniBooNE experimental points are taken from \cite{AguilarArevalo:2010zc}.}
\label{fig_d2s_vs_cos_tmu_025m_035m}
\end{center}
\end{figure}
The bounds for the electron energy are visible in Fig. \ref{fig_cos_rec_emu} in which the abscissa is now the electron energy. 
They correspond to the extreme values of $\cos\theta=\pm 1$. Neglecting the  electron mass the bounds for its energy $E_e$ are
  $\overline{E_\nu}/(1+(2\overline{E_\nu}/M)) < E_e <\overline{E_\nu}$,
 or 0.2 GeV$< E_e <$ 0.3 GeV, which leads for the bounds of the energy transfer,
 0.4 GeV$<\omega= (E_\nu - E_e) <$ 0.5 GeV. The question is then if this region is covered by the data 
on the double differential cross section $\frac{d^2 \sigma}{dE_{\mu}~d\mathrm{cos}\theta}$ measured by MiniBooNE. In this cross section  
as given in Eq. (\ref{cross}), the muon neutrino flux, peaked at $E_{\nu} \simeq 0.7$ GeV, 
favors the region of muon energy in the vicinity of  $E_\mu = 0.7~\textrm{GeV}-\omega$. For  0.4 GeV$<\omega < $0.5 GeV 
the region favored by the peak position is between $E_\mu=$ 0.2 and 0.3 GeV.  As the muon neutrino energy peak is broad, somewhat larger  $E_\mu$ values are relevant as well to test the region of energy transfers between 0.4 and 0.5 GeV. 
We  select therefore for the test of our description of the np-nh component in the region of interest  the two smallest values for $E_\mu$ experimentally  available, $E_\mu=$0.35 GeV and 
$E_\mu=$0.45 GeV. The comparison with the MiniBooNE data is shown in Fig. \ref{fig_d2s_vs_cos_tmu_025m_035m}, reproduced from our previous work \cite{Martini:2011wp}.  The np-nh part is clearly needed to account for the data.

We can therefore conclude that there are experimental indications to confirm the importance of the high-energy np-nh tail and hence the reshaping towards higher energies of the MiniBooNE distribution of the low-energy electron excess when 
expressed in terms of the real neutrino energy rather than of the reconstructed value.

\subsection{T2K} 
\begin{figure}
\begin{center}
  \includegraphics[width=12cm,height=8cm]{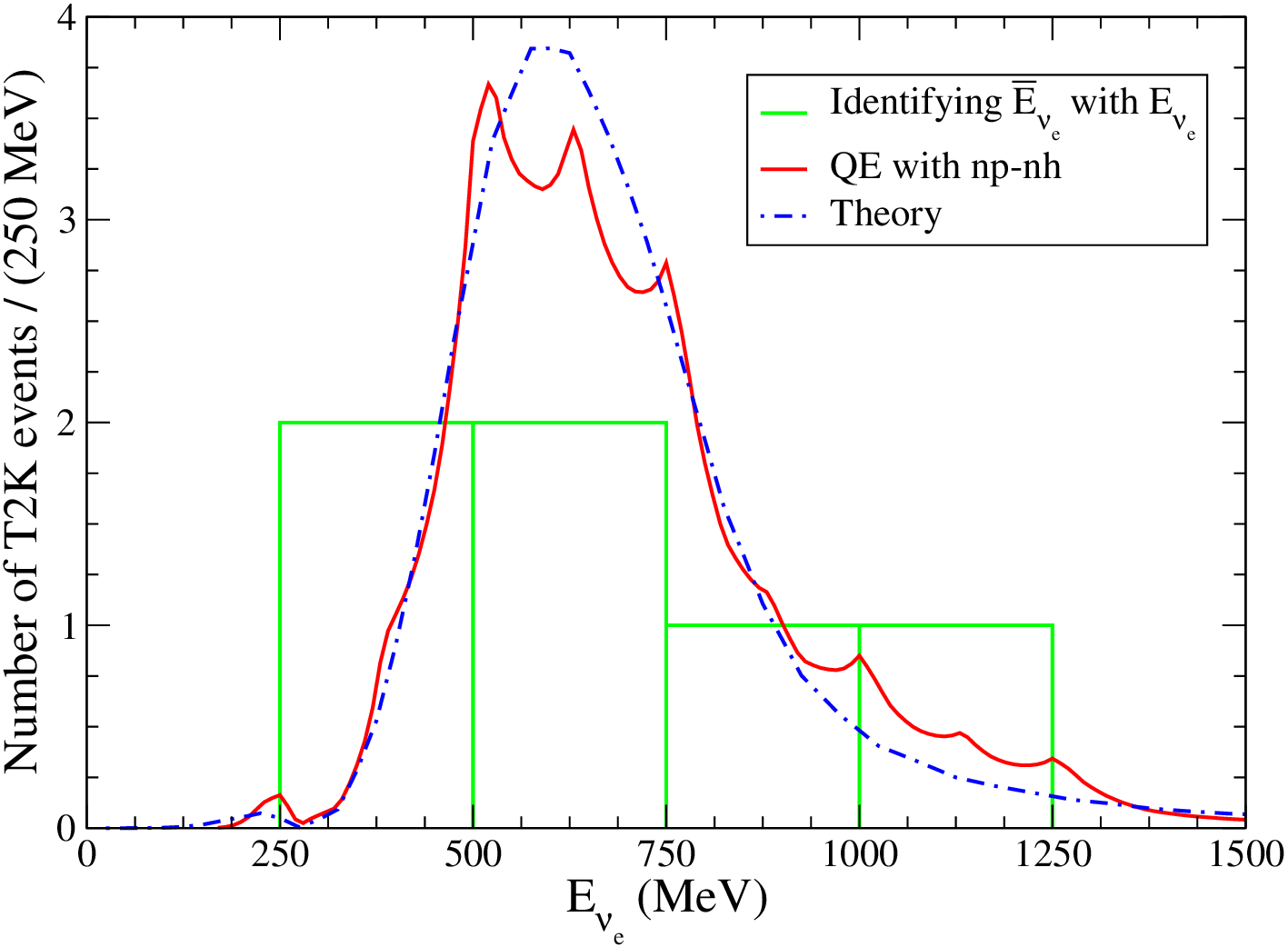}
\caption{(Color online) The experimental T2K data \cite{Abe:2011sj} on electron neutrino excess events. 
The histogram is obtained identifying the reconstructed neutrino energy $\overline{E_\nu}$ with the true neutrino energy $E_\nu$. 
The full line is our smeared distribution including the multinucleon part. 
The dot-dashed line is our theoretical prediction, i.e., the product  
  $\sigma_{\textrm{QE+np-nh}}(E_{\nu_{e}})\Phi_{T2K}^{SK}(E_{\nu_{e}})$ with an area normalized to the total number of events.}
\label{fig_t2k_events_isto}
\end{center}
\end{figure}

The electron neutrino energy distribution at the far Super Kamiokande detector as calculated from the oscillation expression 
has been displayed in Fig. \ref{fig_minib_t2k_flux}. The spectrum is much narrower than the presumed one for the MiniBooNE electron excess events. Hence the reshaping features are somewhat different. On the experimental side the detection of excess electron is still in an early stage and some candidates (6) have been detected. The six electron events are spread on large bins of reconstructed energy. Their energy distribution  is shown in the histogram of Fig. \ref{fig_t2k_events_isto} (assuming  $\overline{E_\nu}\equiv E_\nu$). We also show in this figure our smeared distribution with the same methods as for MiniBooNE. The main feature here is the narrowing of our energy distribution as compared to the experimental histogram which assumes the identity of the reconstructed and the true energy. We have explained the origin of this narrowing from the different behavior of the smearing function at small and large  $\overline{E_\nu}$ values. In the same figure we also show our theoretical prediction for this distribution. Since it represents the interaction number for the electron neutrinos which underwent a ``quasielastic'' event,  it should be given by the quantity 
$\sigma_{\textrm{QE+np-nh}}(E_{\nu_{e}})\Phi_{T2K}^{SK}(E_{\nu_{e}})$, product of the  total ``quasielastic'' cross section by the electron neutrino energy distribution. The area below the theoretical curve has been normalized to the same total number of events. We take for the cross section, $\sigma_{\textrm{QE+np-nh}}(E_{\nu_{e}})$, our calculated value, similar to the one for muon neutrinos \cite{Martini:2009uj} which accounted quite well for the MiniBooNE data. Notice that, due to the normalization to the number of electron events, the introduction of the np-nh component in the ``quasielastic'' cross section has practically no effect, since it amounts essentially to an increase of the cross section by a constant multiplying factor. It is in the narrowing of  the smeared distribution that  the np-nh shows up. In spite of the small number of events it appears that the narrowing goes well in the direction of the theoretical prediction. This offers a consistency test. If the electrons were background ones there would be no reason for an agreement.
A similar comparison will be performed for MiniBooNE when  a better smearing function will be evaluated.
 
\section{Conclusion} 

We have examined in this work the validity of the approximation contained in the 
identification of the reconstructed neutrino energy with the real energy. 
The errors introduced by this assumption have various origins. 
The straightforward ones are the Pauli blocking effect, the Fermi motion which broadens the region of response and the collective aspects of the response. 
However the main correction arises from  the existence of the multinucleon component in the neutrino cross section on nuclei 
which can in Cherenkov detectors simulate a quasielastic process.
We have introduced the probability distribution to have a real neutrino energy $E_{\nu}$, given a reconstructed energy value  $\overline{E_\nu}$. We have found that at low  $\overline{E_\nu}$ values  
this distribution acquires an important high-energy tail due to the multinucleon component. For large reconstructed energies    
it is instead the low-energy tail which is favored. 
  The neutrino energy distribution is an important ingredient in these evaluations since it appears as a weight factor in the cross sections. Its influence can be concisely described as follows: the transition from the reconstructed to the real energy tends to concentrate the events in the region of high flux provided there is some strength in this region.

For the MiniBooNE data on the energy distribution of the excess electron events, with some simplified assumptions on the electron neutrinos energy  distribution, we find that it  is modified with a suppression of the low energy peak, shifted at higher energies. This feature, when confirmed, will  have consequences for the compatibility with other short-baseline data, in particular with the LSND ones.  In this respect one  question will have to be answered:  if other short-baseline experiments should not be affected in a similar way.  There is no general answer and the type of neutrino reaction as well as the detection method should be examined case by case. For example the analysis of MiniBooNE data in the antineutrino mode showing an excess of $\bar{\nu}_e$  events \cite{AguilarArevalo:2010wv,Djurcic:2012jf} would be particularly interesting 
also in connection with CP-violating effect studies. The problem of energy reconstruction in this case is the object of our present investigations.

A reshaping occurs as well for the T2K electron events in the Super Kamiokande detector. In this case the narrowness of the  electron neutrino  energy distribution shows up as a narrowing of the electron events  distribution, which goes in the direction of our theoretical prediction. In a forthcoming work we will also investigate the corresponding problems for the muonic events both in the close and in the far detector. 
 In the latter case the oscillations lead to an energy distribution of the muon neutrino spectrum which is quite different from the one at the close detector \cite{Abe:2012gx}. The reshaping in terms of the true neutrino energy may then reveal some amusing and unexpected features. But we also keep in mind the problems associated with the MiniBooNE data and what we can say at the nuclear light on the challenging  problems of sterile neutrinos and short-baseline oscillations.\\

\noindent{\bf \large{Acknowledgments}}

\noindent
We thank Gerald Garvey who stimulated our interest in the problem of the energy reconstruction. 
We thank Torleif Ericson for useful discussions. This work was partially supported  by the Communaut\'e
Fran\c caise de Belgique (Actions de Recherche Concert\'ees).

\end{document}